\documentclass[twocolumn]{aastex61}
\usepackage{graphicx,mathtools}
\usepackage{amsmath,amssymb}
\usepackage{xparse}
\usepackage{xspace}
\usepackage{natbib}
\usepackage{ulem}

\makeatletter
\def\Ginclude@eps#1{%
 \message{<#1>}%
  \bgroup
  \def\@tempa{!}%
  \dimen@\Gin@req@width
  \dimen@ii.1bp%
  \divide\dimen@\dimen@ii
  \@tempdima\Gin@req@height
  \divide\@tempdima\dimen@ii
    \includegraphics{#1}%
  \egroup}
\makeatother

\DeclareMathOperator{\dd}{{\rm d}}
\DeclareMathOperator{\id}{{\rm d}\!}

\newcommand{\e}[1]{\!\times\! 10^{#1}}

\newcommand{\have}[1]{\left\langle #1 \right\rangle}

\newcommand{\bcave}[1]{\left\lbrace #1 \right\rbrace_t}

\newcommand{\Tep}{\ensuremath{T_{\rm eff}^+}\xspace}
\newcommand{\Tem}{\ensuremath{T_{\rm eff}^-}\xspace}

\begin{document}

\title{Convection Enhances Magnetic Turbulence in AM CVn Accretion Disks}
\shorttitle{Convection Enhances MRI Turbulence in AM CVn Accretion Disks}
\shortauthors{M. S. B. Coleman et al.}

\author{Matthew S. B. Coleman}
\affiliation{School of Natural Sciences, Institute for Advanced Study, Einstein Drive, Princeton, NJ 08540, USA; {\rm \url{mcoleman@ias.edu}}}
\affiliation{Department of Physics, University of California, Santa Barbara, CA
93106, USA}

\author{Omer Blaes}
\affiliation{Department of Physics, University of California, Santa Barbara, CA
93106, USA}

\author{Shigenobu Hirose}
\affiliation{Department of Mathematical Science and Advanced Technology, Japan
Agency for Marine-Earth Science and Technology, Yokohama, Kanagawa 236-0001, Japan}

\author{Peter H. Hauschildt}
\affiliation{Hamburger Sternwarte, Gojenbergsweg 112, D-21029 Hamburg, Germany}

\begin{abstract}
We present the results of local, vertically stratified, radiation magnetohydrodynamic shearing box simulations of magnetorotational instability (MRI) turbulence for a (hydrogen poor) composition applicable to accretion disks in AM CVn type systems. 
Many of these accreting white dwarf systems are helium analogues of dwarf novae (DNe).
We utilize frequency-integrated opacity and equation of state tables appropriate for this regime to accurately portray the relevant thermodynamics.
We find bistability of thermal equilibria in the effective temperature, surface mass density plane typically associated with disk instabilities.
Along this equilibrium curve (i.e. the S-curve) we find that the stress to thermal pressure ratio $\alpha$ varied with peak values of $\sim 0.15$ near the tip of the upper branch. 
Similar to DNe, we found enhancement of $\alpha$ near the tip of the upper branch caused by convection; this increase in $\alpha$ occurred despite our choice of zero net vertical magnetic flux.
Two notable differences we find between DN and AM~CVn accretion disk simulations are that AM~CVn disks are capable of exhibiting persistent convection in outburst, and ideal MHD is valid throughout quiescence for AM~CVns. In contrast, DNe simulations only show intermittent convection, and non-ideal MHD effects are likely important in quiescence. By combining our previous work with these new results, we also find that convective enhancement of the MRI is anticorrelated with mean molecular weight.
\end{abstract}

\section{Introduction}

AM Canum Venaticorum type stars (AM~CVns) are very short period ($\lesssim 65$ minutes) binary systems consisting of a white dwarf that is accreting from a
Roche lobe filling, hydrogen-deficient companion star (often another, lower mass, white dwarf), and can be thought of
as helium analogues of cataclysmic variables (see, e.g.,
\citealt{SOL10} for a review).  They are of intrinsic astrophysical interest
because they may be sources of helium novae, .Ia supernovae
\citep{BIL07}, and possibly even type Ia supernovae, as well as being among
the strongest known
sources of gravitational waves detectable by the future space-based gravitational
wave mission LISA (e.g. \citealt{NEL04}).
The helium-dominated accretion disks in AM~CVns exhibit all the rich
phenomenology of hydrogen-dominated
accretion disks around white dwarfs, including
persistent and outbursting systems, eclipsing systems, superhumps,
quasiperiodic oscillations (QPO’s),
and broadband noise (e.g. \citealt{RAM12,CAM15,KUP15,LEV15}).
The outbursting systems are generally
dominated by superoutbursts, although normal outbursts have also been observed
\citep{LEV11}.

Ionization driven thermal instability models of outbursting helium accretion
disks were first considered by \citet{SMAK83} and \citet{CAN84}, and more recent
models have been developed by \citet{KOT12} and \citet{CAN15}.
As in the standard disk instability models of hydrogen-rich dwarf
novae (DNe), these helium dominated models require that the \citet{SS73}
$\alpha$ parameter be
larger in outburst than in quiescence, though the enhancement of $\alpha$ need not
be as large (possibly as low as a factor of two) as compared to hydrogen
accreting systems \citep{KOT12}.
The value of $\alpha$ in outburst in AM~CVn systems
appears to be similar to the outburst value of DNe (0.1-0.2),
though the observational constraints are currently weaker than in DNe
\citep{KOT12, KL12}.

Explaining the observationally-inferred variation in $\alpha$ across the
ionization transition has been a challenge.  One possibility is that it is
related to a high resistivity in the quiescent (and predominantly electrically
neutral) state, which would hinder MHD turbulence and reduce $\alpha$ there
\citep{GAM98}.  However, in the outburst state, where ideal MHD should hold,
simulations of magnetorotational (MRI, \citealt{BAL91, BAL98}) turbulence
with no externally imposed vertical
magnetic flux typically give time-averaged $\alpha$ values of only a few
percent.\footnote{Questions of numerical convergence still exist for simulations
which lack vertical magnetic flux and thermodynamics, e.g. \citet{RYA17}.
Also, there may be
a dependence of $\alpha$ on box height in local shearing box simulations
\citep{SHI16}.}
Including a net vertical magnetic field can enhance the
value of $\alpha$ \citep{HAW95, SAN04, PES07},
but then this begs the question as to why the outburst $\alpha$ values
should be so similar across sources.

On the other hand, incorporating the thermodynamics of radiative cooling
with realistic opacities and ionizing equation of state in stratified
shearing box simulations, \cite{HIR14}
found that thermal convection is driven near the ionization transition.
This modifies the MRI turbulence so as to enhance $\alpha$ in outburst just
above the transition to quiescence.  Values of $\alpha$ there were found to be
as high as 0.14, even without the presence of net vertical magnetic flux.
This convection-driven enhancement of $\alpha$ has since been confirmed
using independent numerical codes \citep{SCE18}, and
was also found just above the
hydrogen ionization transition under conditions relevant for the inner
regions of protoplanetary disks \citep{HIR15}.  In addition to the enhancement
of turbulent stresses, convection also alters the character of the MRI
``butterfly diagram" dynamo, quenching field reversals because of inward
advection of magnetic field with consistent polarity from high altitude
\citep{COL17}.

The purpose of this paper is to extend this work on the effects of convection
on MRI turbulent stresses to AM~CVn disks, which have very different chemical
compositions (dominated by helium) to those in DNe (dominated by hydrogen).
There are potentially two interesting effects:  (1) the existence of two
ionization stages of helium may affect the properties of convection,
and (2) the high first ionization potential
of HeI means that there will be many more free electrons from ionized carbon,
nitrogen and oxygen in the quiescent state compared to that in a hydrogen
disk, thereby enhancing the electrical conductivity.

The structure of this paper is as follows.  In section 2 we briefly review
our computational methods, opacities and equation of state, and simulation
parameters.  In section 3 we discuss our results on the thermal bistability
of a helium disk, the MRI turbulent stresses, the properties of convection,
and the effects on the MRI dynamo.  We discuss the implications of our
results in section 4, and summarize our conclusions in section 5.

\begin{deluxetable}{cccc}
	\tablecaption{Elemental Abundances\label{table:abund}}
	\tablehead{
		\colhead{Element} & \colhead{Log num frac} & \colhead{Log mass frac} & \colhead{Source}
	}
	\startdata
	H & -17.1 & -17.7 & AGS05$/10^{18}$\\
	He & -0.00383 & -0.0157 & BBMP15\\
	C & -2.51 & -2.05 & BBMP15\\
	N & -2.45 & -1.92 & BBMP15\\
	O & -3.67 & -3.08 & BBMP15\\
	Ne & -3.26 & -2.57 & BBMP15\\
	Na & -4.88 & -4.14 & AGS05\\
	Mg & -3.16 & -2.39 & BBMP15\\
	Al & -4.68 & -3.87 & AGS05\\
	Si & -3.54 & -2.71 & AGS05\\
	S & -3.91 & -3.02 & AGS05\\
	Ar & -4.87 & -3.89 & AGS05\\
	Ca & -4.74 & -3.76 & AGS05\\
	Fe & -3.60 & -2.47 & AGS05\\
	Ni & -4.82 & -3.67 & AGS05\\
	\enddata
	\tablecomments{
		Partial list of elemental abundances assumed in this work listed in terms of log$_{10}$ of number fraction and log$_{10}$ of mass fraction. Abundances with the source BBMP15 are computed from the mass fraction of the predominant isotope in the accreted matter in \citet{BRO15}. Abundances from AGS05 are solar abundances as reported by \citet{AGS05} such that the relative abundance between a given element and the combination of C, N, and O is preserved. For hydrogen we assumed that its number fraction is $10^{-18}$ times that in \citet{AGS05}. For brevity, elements (other than H) with number fractions below $10^{-5}$ are not listed here but are assumed to be consistent with AGS05.}
\end{deluxetable}

\begin{figure}
	\includegraphics[width=\linewidth]{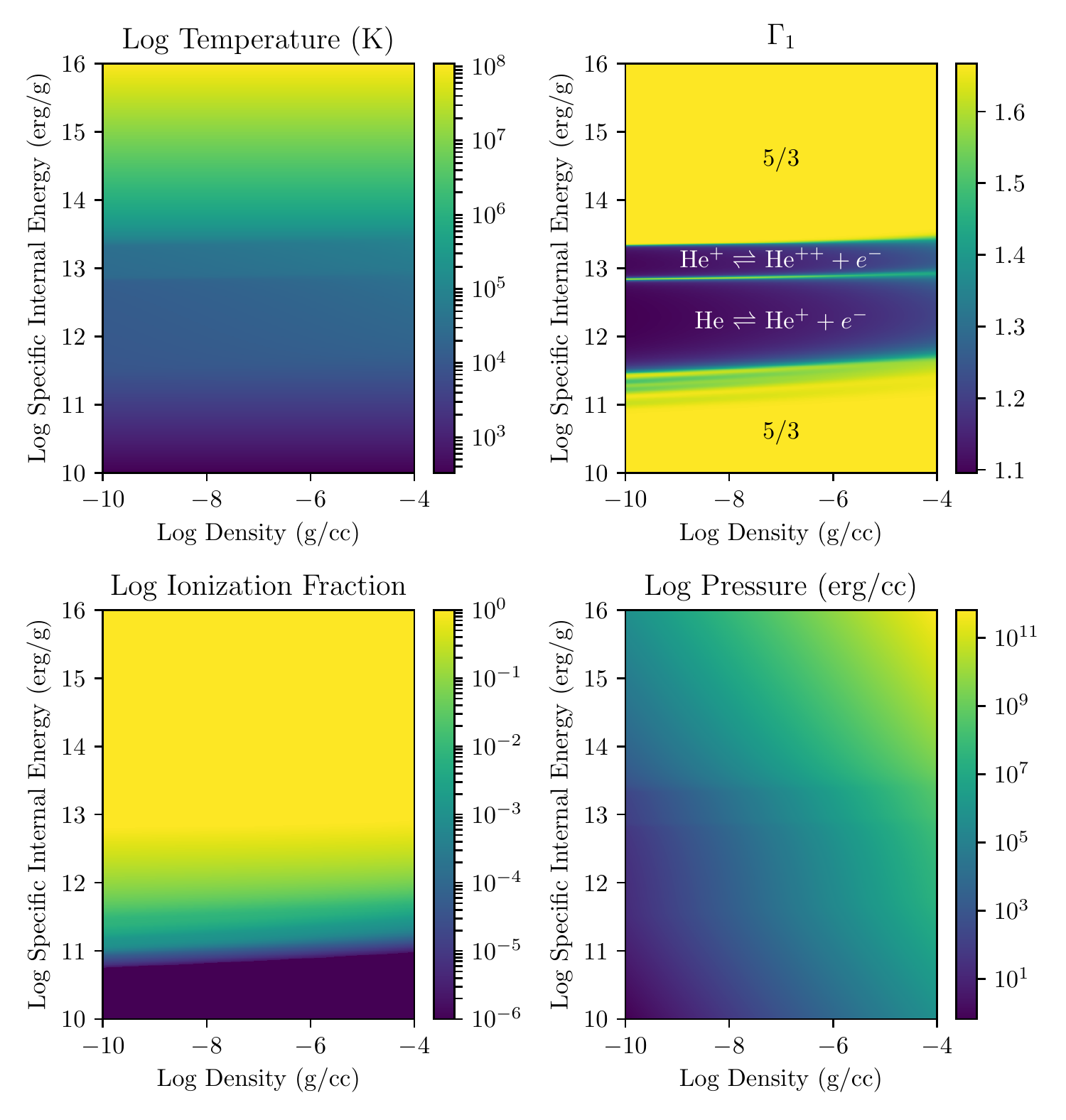}
	\caption{Various equation of state (EOS) parameters as functions of density ($\rho$) and specific internal energy ($e/\rho$). From left to right, top to bottom:
		gas temperature,
		generalized adiabatic index $\Gamma_1\equiv\left(\partial\ln P/\partial\ln\rho\right)_s$,
		ionization fraction,
		and gas pressure.
		For $\Gamma_1$ we have denoted the ionization transitions of He and the asymptotic $\Gamma_1=5/3$ limits.
		Here, we have defined the ionization fraction to be the fraction of atoms which are at least singly ionized, explaining why our values asymptote to unity at high temperature.
	}
	\label{fig:eos}
\end{figure}

\begin{figure}
	\includegraphics[width=1.0\linewidth]{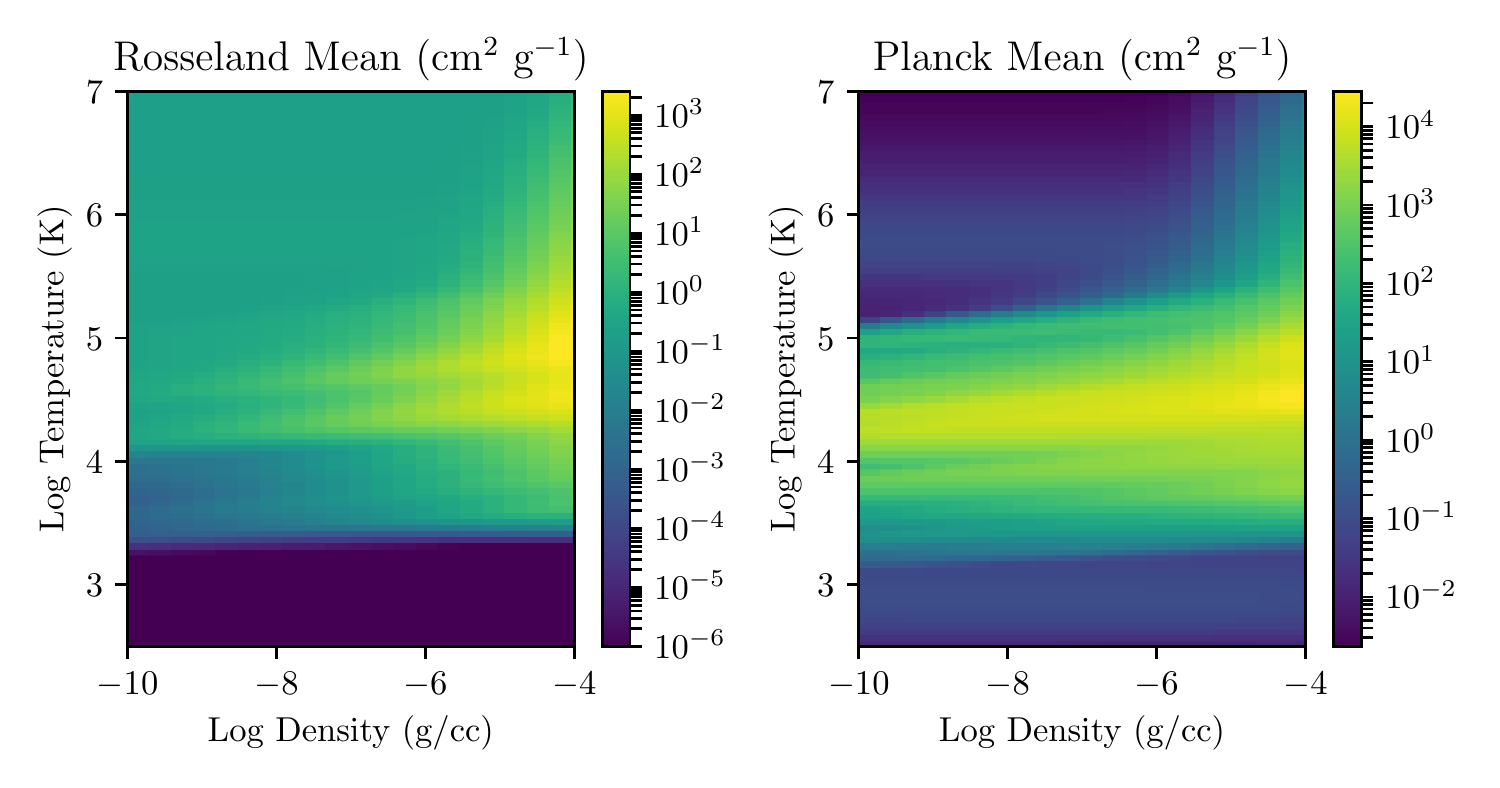}
	\caption{Rosseland and Planck mean dust-free opacity tables as functions of density and temperature. Note the different ranges on the color scales, indicating that the Rosseland mean opacity exhibits significantly more variation compared to the Planck mean.
	}
	\label{fig:opacity}
\end{figure}

\section{Methods}

We utilize the {\sc zeus} code \citep{Zeus1, Zeus2} with flux limited diffusion \citep{ZeusRad} to evolve the equations of radiation magnetohydrodynamics. Our zero net vertical flux shearing-box simulations are based on those presented in \citet{HIR14}, to which we refer the reader for a more thorough discussion of the equations and numerical techniques.  The primary difference with that earlier work is composition. Here we assume that hydrogen is negligible and helium is the dominant species (see Table \ref{table:abund}), necessitating new equation of state (EOS) and opacity tables.

\subsection{Composition}

The lack of spectral hydrogen signatures in AM~CVns \citep[e.g.][]{KUP11} signifies a significant departure from solar composition. In addition to this dramatic lack of hydrogen, more subtle differences in the composition of AM~CVns result from CNO burning in the
prior evolution of the donor star \citep{SOL10}. We therefore use a model of an AM~CVn from \citet{BRO15} as a starting point for our assumed chemical abundances. We examined the freshly accreted material on the surface of the primary white dwarf in this model and used the measured mass fraction of the predominant isotope of several elements as the basis of our composition.

For the remaining elements we assume that the CNO burning which took place in the prior evolutionary states of AM~CVns preserves the combined number of C, N and O and leaves the number fraction of most other elements unchanged. For elements not taken from \citet{BRO15}, we assumed that their abundance relative to the combination of C, N and O is identical to solar composition \citep{AGS05}. This means that our assumed composition is consistent with that of a binary star system which started with solar metallicity and evolved into an AM~CVn.

These changes in composition (as well as the orbital frequency $\Omega$) are the primary differences from the {\sc zeus} simulations presented in \citet{HIR14, HIR15, COL16}, and are only manifest in the EOS and opacity tables. The EOS framework \citep[see Appendix A of][for details]{diss} used in the shearing box simulations is the same as before, now run with the abundances listed in Table \ref{table:abund}, the results of which are shown in Figure~\ref{fig:eos}. We also used this composition to compute our own Rosseland and Planck mean opacities with the latest version of the stellar atmosphere code {\sc phoenix} \citep{FER05}, with dust explicitly turned off (both in the EOS and opacities). These dust-free tables are shown in Figure~\ref{fig:opacity}. We also note that {\sc phoenix} computes its own EOS which is consistent with that used in our {\sc zeus} simulations but utilizes a more detailed calculation, required to accurately compute the opacities.

\begin{longrotatetable}
	\begin{deluxetable}{cccccccccccccccccccccc}
		\tablecaption{Simulation parameters\label{table:param}}
		\renewcommand{\arraystretch}{.8}
		\tabletypesize{\small}
		\tablehead{
			\colhead{Run} & \colhead{$\Sigma_0$} & \colhead{$T_{\rm eff,0}$} & \colhead{$\Sigma$} & \colhead{$T_{\rm eff}$} & \colhead{$T_{\rm c}$} & \colhead{$\tau_{\rm tot}$} & \colhead{$\alpha$} & \colhead{$f_{\rm adv}$} & \colhead{$10^2M_{\rm adv}$} & \colhead{$t_{\rm th}$} & \colhead{$\frac{h_0}{10^8}$} & \colhead{$\frac{h_P}{h_0}$} & \colhead{$\frac{h_{\rm phot}}{h_0}$} & \colhead{$N_x$} & \colhead{$N_y$} & \colhead{$N_z$} & \colhead{$\frac{L_x}{h_0}\!$} & \colhead{$\frac{L_y}{h_0}\!$} & \colhead{$\frac{L_z}{h_0}\!$} & \colhead{$t_1$} & \colhead{$t_2$}
		}
		\startdata
		$\Sigma$5.3e3-U0 & 5332 & 40000 & 5287 & 41880 & 4.55e5 & 136106 & 0.016 & $\sim\!0$ & $\sim\!0$ & 20 & 1.07 & 0.545 & 1.87 & 32 & 64 & 288 & 1 & 2 & 4.5 & 23 & 123\\
		$\Sigma$4.4e3-U0 & 4356 & 40000 & 4309 & 40346 & 4.13e5 & 108428 & 0.019 & $\sim\!0$ & $\sim\!0$ & 19 & 1.02 & 0.542 & 1.87 & 32 & 64 & 288 & 1 & 2 & 4.5 & 44 & 144\\
		$\Sigma$1.9e3-U0 & 1915 & 33000 & 1900 & 29203 & 2.66e5 & 61612 & 0.019 & $\sim\!0$ & $\sim\!0$ & 21 & 0.803 & 0.546 & 1.71 & 32 & 64 & 288 & 1 & 2 & 4.5 & 22 & 122\\
		$\Sigma$9.6e2-U0 & 956 & 25000 & 946 & 24082 & 1.98e5 & 37681 & 0.024 & $\sim\!0$ & $\sim\!0$ & 18 & 0.668 & 0.56 & 1.66 & 32 & 64 & 288 & 1 & 2 & 4.5 & 21 & 121\\
		$\Sigma$4.7e2-U0 & 475 & 20000 & 466 & 18655 & 1.34e5 & 24030 & 0.026 & $\sim\!0$ & 0.065 & 20 & 0.543 & 0.563 & 1.49 & 32 & 64 & 288 & 1 & 2 & 4.5 & 42 & 142\\
		$\Sigma$3.7e2-U0 & 375 & 17500 & 366 & 17083 & 1.11e5 & 30662 & 0.030 & 0.228 & 0.301 & 20 & 0.478 & 0.572 & 1.53 & 32 & 64 & 288 & 1 & 2 & 4.5 & 37 & 137\\
		$\Sigma$3.3e2-U2 & 331 & 20000 & 324 & 17749 & 1.14e5 & 19491 & 0.037 & 0.087 & 0.207 & 16 & 0.502 & 0.558 & 1.53 & 32 & 64 & 288 & 1 & 2 & 4.5 & 30 & 130\\
		$\Sigma$2.8e2-U0 & 283 & 20000 & 277 & 16643 & 9.96e4 & 25966 & 0.040 & 0.314 & 0.473 & 16 & 0.486 & 0.529 & 1.4 & 32 & 64 & 288 & 1 & 2 & 4.5 & 45 & 145\\
		$\Sigma$2.3e2-U0 & 232 & 17500 & 226 & 16001 & 9.03e4 & 23640 & 0.047 & 0.395 & 0.639 & 14 & 0.431 & 0.562 & 1.45 & 32 & 64 & 288 & 1 & 2 & 4.5 & 31 & 131\\
		$\Sigma$1.8e2-U1 & 184 & 17000 & 176 & 15366 & 8.12e4 & 19955 & 0.057 & 0.528 & 0.915 & 12 & 0.396 & 0.566 & 1.35 & 32 & 64 & 288 & 1 & 2 & 4.5 & 70 & 170\\
		$\Sigma$1.6e2-U0 & 165 & 17000 & 155 & 14943 & 7.45e4 & 19540 & 0.066 & 0.605 & 1.12 & 10 & 0.385 & 0.549 & 1.25 & 32 & 64 & 288 & 1 & 2 & 4.5 & 80 & 180\\
		$\Sigma$1.6e2-U1 & 156 & 17000 & 149 & 14407 & 6.66e4 & 22414 & 0.070 & 0.724 & 1.37 & 9.83 & 0.379 & 0.514 & 1.22 & 32 & 64 & 288 & 1 & 2 & 4.5 & 79 & 179\\
		$\Sigma$1.5e2-U0 & 153 & 16000 & 145 & 13401 & 5.63e4 & 23304 & 0.073 & 0.804 & 1.71 & 8.73 & 0.274 & 0.62 & 1.24 & 32 & 64 & 288 & 1 & 2 & 4.5 & 33 & 133\\
		$\Sigma$1.4e2-U2 & 137 & 17000 & 133 & 14402 & 6.52e4 & 18694 & 0.079 & 0.726 & 1.6 & 8.83 & 0.364 & 0.532 & 1.09 & 32 & 64 & 288 & 1 & 2 & 4.5 & 30 & 130\\
		$\Sigma$1.3e2-U1 & 128 & 17000 & 120 & 13677 & 5.77e4 & 15918 & 0.091 & 0.747 & 1.8 & 7.57 & 0.355 & 0.485 & 1.02 & 32 & 64 & 288 & 1 & 2 & 4.5 & 80 & 180\\
		$\Sigma$1.2e2-U1 & 118 & 17000 & 110 & 13732 & 5.82e4 & 14032 & 0.098 & 0.715 & 1.85 & 7.16 & 0.343 & 0.514 & 0.992 & 32 & 64 & 288 & 1 & 2 & 4.5 & 80 & 180\\
		$\Sigma$1.1e2-U1 & 109 & 17000 & 101 & 12028 & 4.11e4 & 13080 & 0.113 & 0.768 & 2.25 & 6.03 & 0.331 & 0.388 & 0.82 & 32 & 64 & 288 & 1 & 2 & 4.5 & 80 & 180\\
		$\Sigma$8.9e1-U1 & 89 & 16500 & 85 & 11599 & 3.74e4 & 9517 & 0.146 & 0.751 & 2.82 & 4.76 & 0.264 & 0.453 & 0.883 & 32 & 64 & 288 & 1 & 2 & 4.5 & 32 & 132\\
		\hline
		$\Sigma$1.9e2-M0 & 192 & 7400 & 180 & 8976 & 2.48e4 & 40760 & 0.047 & 0.884 & 1.36 & 11 & 0.111 & 0.79 & 1.48 & 32 & 64 & 288 & 1 & 2 & 4.5 & 26 & 126\\
		$\Sigma$2.0e2-M0 & 201 & 7400 & 193 & 8676 & 2.40e4 & 40225 & 0.041 & 0.881 & 1.23 & 12 & 0.113 & 0.751 & 1.44 & 32 & 64 & 288 & 1 & 2 & 4.5 & 27 & 127\\
		$\Sigma$2.0e2-M1 & 201 & 7000 & 192 & 8632 & 2.40e4 & 39851 & 0.041 & 0.881 & 1.3 & 12 & 0.11 & 0.773 & 1.44 & 32 & 64 & 288 & 1 & 2 & 4.5 & 33 & 133\\
		\hline
		$\Sigma$2.1e2-L1 & 210 & 6000 & 202 & 7651 & 2.21e4 & 28935 & 0.028 & 0.827 & 0.906 & 14 & 0.103 & 0.751 & 1.37 & 32 & 64 & 288 & 1 & 2 & 4.5 & 42 & 142\\
		$\Sigma$1.9e2-L1 & 192 & 7000 & 191 & 6129 & 1.90e4 & 9148 & 0.014 & 0.675 & 0.508 & 18 & 0.108 & 0.626 & 1.08 & 32 & 64 & 288 & 1 & 2 & 4.5 & 46 & 146\\
		$\Sigma$1.8e2-L0 & 182 & 16000 & 179 & 6889 & 2.00e4 & 12897 & 0.022 & 0.727 & 0.741 & 15 & 0.122 & 0.582 & 1.04 & 32 & 64 & 288 & 1 & 2 & 4.5 & 62 & 162\\
		$\Sigma$1.7e2-L0 & 173 & 7500 & 172 & 6054 & 1.84e4 & 6225 & 0.016 & 0.591 & 0.455 & 15 & 0.108 & 0.607 & 1.04 & 32 & 64 & 288 & 1 & 2 & 4.5 & 79 & 179\\
		$\Sigma$1.3e2-L0 & 130 & 10000 & 128 & 5017 & 1.36e4 & 865 & 0.016 & 0.208 & 0.156 & 14 & 0.119 & 0.451 & 0.703 & 32 & 64 & 288 & 1 & 2 & 4.5 & 60 & 160\\
		$\Sigma$1.3e2-L1 & 129 & 8700 & 128 & 4390 & 1.05e4 & 599 & 0.012 & 0.243 & 0.141 & 20 & 0.108 & 0.434 & 0.664 & 32 & 64 & 288 & 1 & 2 & 4.5 & 122 & 222\\
		$\Sigma$1.3e2-L2 & 129 & 6000 & 126 & 5169 & 1.49e4 & 1111 & 0.015 & 0.239 & 0.189 & 15 & 0.089 & 0.632 & 1.01 & 32 & 64 & 288 & 1 & 2 & 4.5 & 38 & 138\\
		$\Sigma$1.0e2-L1 & 100 & 5000 & 98 & 4152 & 8.88e3 & 286 & 0.016 & 0.182 & 0.144 & 13 & 0.070 & 0.608 & 0.867 & 32 & 64 & 288 & 1 & 2 & 4.5 & 35 & 135\\
		$\Sigma$7.5e1-L2 & 75 & 3500 & 70 & 3088 & 2.18e3 & 4e-3 & 0.022 & $\sim\!0$ & $\sim\!0$ & 9.31 & 0.041 & 0.606 & N/A & 32 & 64 & 288 & 1 & 2 & 4.5 & 80 & 180\\
		\enddata
		\tablecomments{
			Each row corresponds to the parameters of a simulation where the name of the run is specified in the first column. The second to last character of a run name denotes the branch which the simulation settled to: ``U" for upper, ``M" for middle, and ``L" for lower.
			Times ($t_{\rm th,1,2}$) are listed in orbital periods, all other numbers are given in cgs units. Zero in the subscript denotes initial conditions. All other values have been averaged over the time interval from $t_1$ to $t_2$. $\Sigma$ is the surface mass density, $T_{\rm eff,c}$ are the effective and central temperatures, $\tau_{\rm tot}$ is the total optical depth integrated between the vertical bounds of the simulation, $\alpha$ is the ratio of volume and time-averaged stressed to volume and time-averaged thermal pressure, $f_{\rm adv}$ and $M_{\rm adv}$ are estimators of convective strength and are given by Eqns. \ref{eqn:fadv} and \ref{eqn:Madv} respectively, $t_{\rm th}$ is the thermal time, $h_0$ is the simulation length unit, $h_P$ is the pressure scale-height, $h_{\rm phot}$ is the photosphere height, $N_{x,y,x}$ are the number of cells in the coordinate directions, and $L_{x,y,z}$ are the lengths of the simulation domain.
		}
	\end{deluxetable}
\end{longrotatetable}

\begin{figure*}
	\hfill\includegraphics[width=\linewidth]{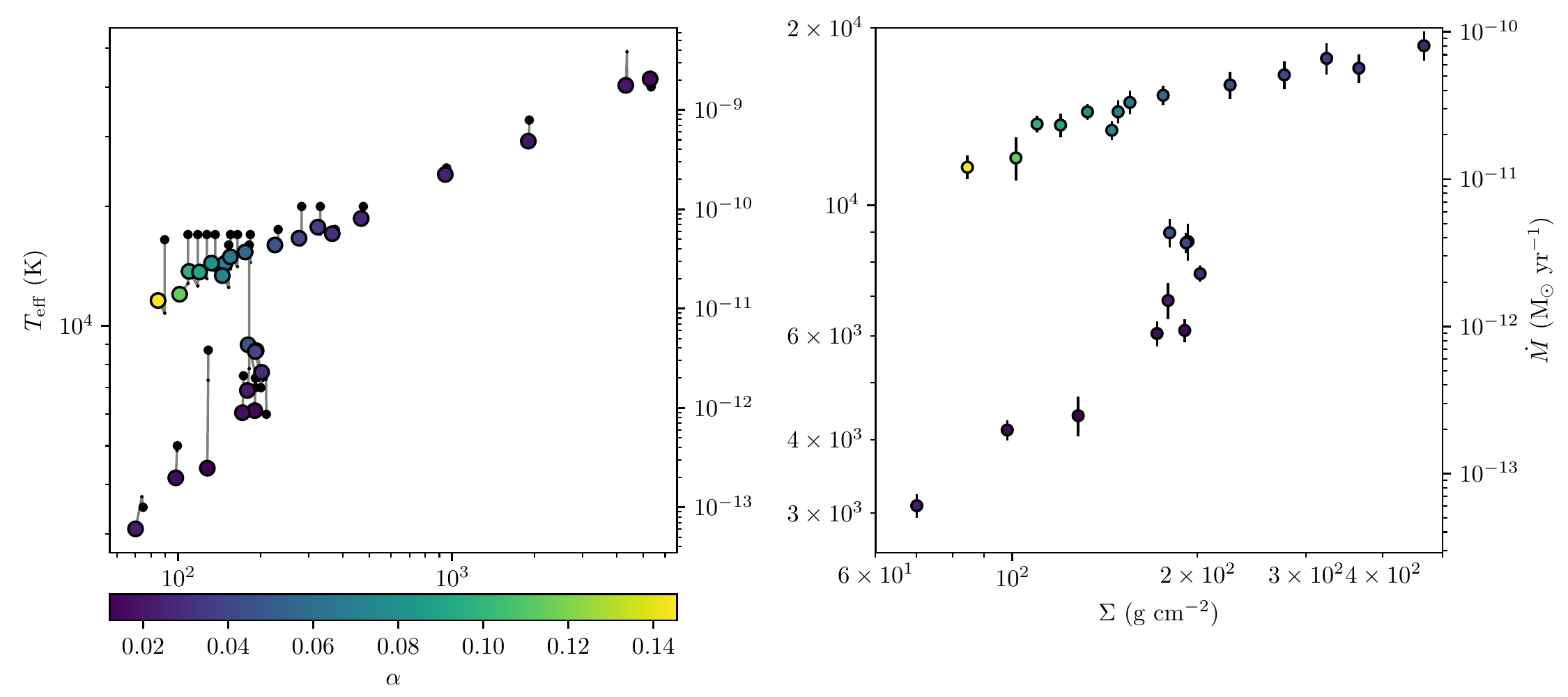}
	\caption{Thermal equilibria achieved by simulations plotted in surface mass density ($\Sigma$) effective temperature ($T_{\rm eff}$) space. This effective temperature also corresponds to a mass accretion rate via Eqn.~\ref{eqn:balance} which is shown on the right axis of both panels. Each colored point represents the time averaged quantities of a single simulation where the color denotes the time averaged $\alpha$ value. In the left panel the large black dot signifies the initial conditions and the small black dot is the data after 10 orbits. Gray lines connects these three points for each simulation. The right panel zooms in on the data, and here we have included vertical black bars whose hight correspond to $\pm$ one standard deviation of the temporal variation (from $t_1$ to $t_2$ in Table~\ref{table:param}) of the effective temperature for each simulation.
	}
	\label{fig:scurve}
\end{figure*}

\subsection{Parameters}

One of the most important and fundamental parameters for a shearing box is its orbital frequency

\begin{equation}
\Omega = 0.106409\text{ rad s}^{-1}\times\left(\dfrac{r}{5\, R_{\rm WD}}\right)^{-3/2}\left( \dfrac{M_1}{1.1\,M_{\sun}}\right)^{1/2},
\end{equation}
with $R_{\rm WD}=4.68\e{8}$ cm and $M_1=1.1\,M_\sun$ as the assumed radius and mass of the white dwarf primary. We take $\Omega= 0.106409$~rad s$^{-1}$ for all the simulations presented in this work.
The surface mass density $\Sigma=\int \rho\,{\rm d}z$ is also important, as it determines the types of possible thermal equilibria.
Additionally, the net vertical magnetic flux is also an inherently interesting parameter \citep[see e.g.][]{SAL16}, however we keep this zero to minimize our parameter space and to explore enhancements of $\alpha$ where it typically takes its lowest value.
Neglecting the usually small mass loss through our vertical outflow boundary conditions, these three parameters are conserved through the evolution of a shearing box simulation.

In addition to these
conserved parameters the choice of an initial effective temperature is necessary to establish initial conditions based on simplistic hydrostatic and thermal equilibrium \citep[see Section 2.4 of][for more details]{HIR14}. As the simulation evolves the MRI sets in and develops turbulence, typically within 10 orbits. This turbulence naturally results in dissipation occurring throughout the disk, which responds by either heating or cooling until a quasi-steady state is achieved, which may be significantly far from the assumed initial condition. The parameters which define these initial conditions along with several time averaged quantities of our various simulations are shown in Table~\ref{table:param}.

\section{Results}

To discuss our results, it is useful to define the following quantities related to some fluid variable $f$:  the horizontal average of this quantity, 
a one orbit temporally smoothed quantity, the temporal mean, and the mean midplane value. These are
defined respectively by
\begin{subequations}
	\begin{align}
	\have{f}(t,z) &\equiv\frac{1}{L_xL_y}\int_{-L_x/2}^{L_x/2}dx\int_{-L_y/2}^{L_y/2}dy
	f(t,\vec{r})\\
	\bcave{f}(t) &\equiv\left.\int_{t-1/2}^{t+1/2} f(t^\prime) \, dt^\prime \middle/ 1\text{ orbit}\right.\\
	\bar{f}&\equiv\dfrac{1}{t_2-t_1}\int_{t_1}^{t_2}f(t^\prime) \, \id t\\
	f_c(t)&\equiv\have{f}(t,0)
	\end{align}
\end{subequations}
Here $L_x$, $L_y$, and $L_z$ are the radial, azimuthal and vertical extents of the simulation domain, respectively, and $t_1$, $t_2$ are the endpoints used for temporal integration (listed in Table \ref{table:param}).

\subsection{Thermal Equilibria}

An essential aspect of analyzing disk instabilities is understanding the conditions leading to thermal equilibrium.
It is common to depict the thermal equilibria of a local patch of an accretion
disk as a curve in the plane of surface mass density $\Sigma$ and effective
temperature $T_{\rm eff}$.  Near an ionization transition (or other instabilities), this curve typically
has a distinctive shape leading to the name S-curve.
The thermal equilibria resulting from shearing box simulations such as
ours (see Figure~\ref{fig:scurve}) are often missing most, if not all, of the middle branch which gives the curve its distinctive S shape. This is because the negative slope of this region of the curve indicates that these equilibria are formally unstable. However, numerical effects can lead to additional stability in this region, resulting in simulations being erroneously attracted to this
branch (see e.g. Section 5.2 of \citealt{JIA13}).

\begin{figure*}
	\includegraphics[width=1.0\linewidth]{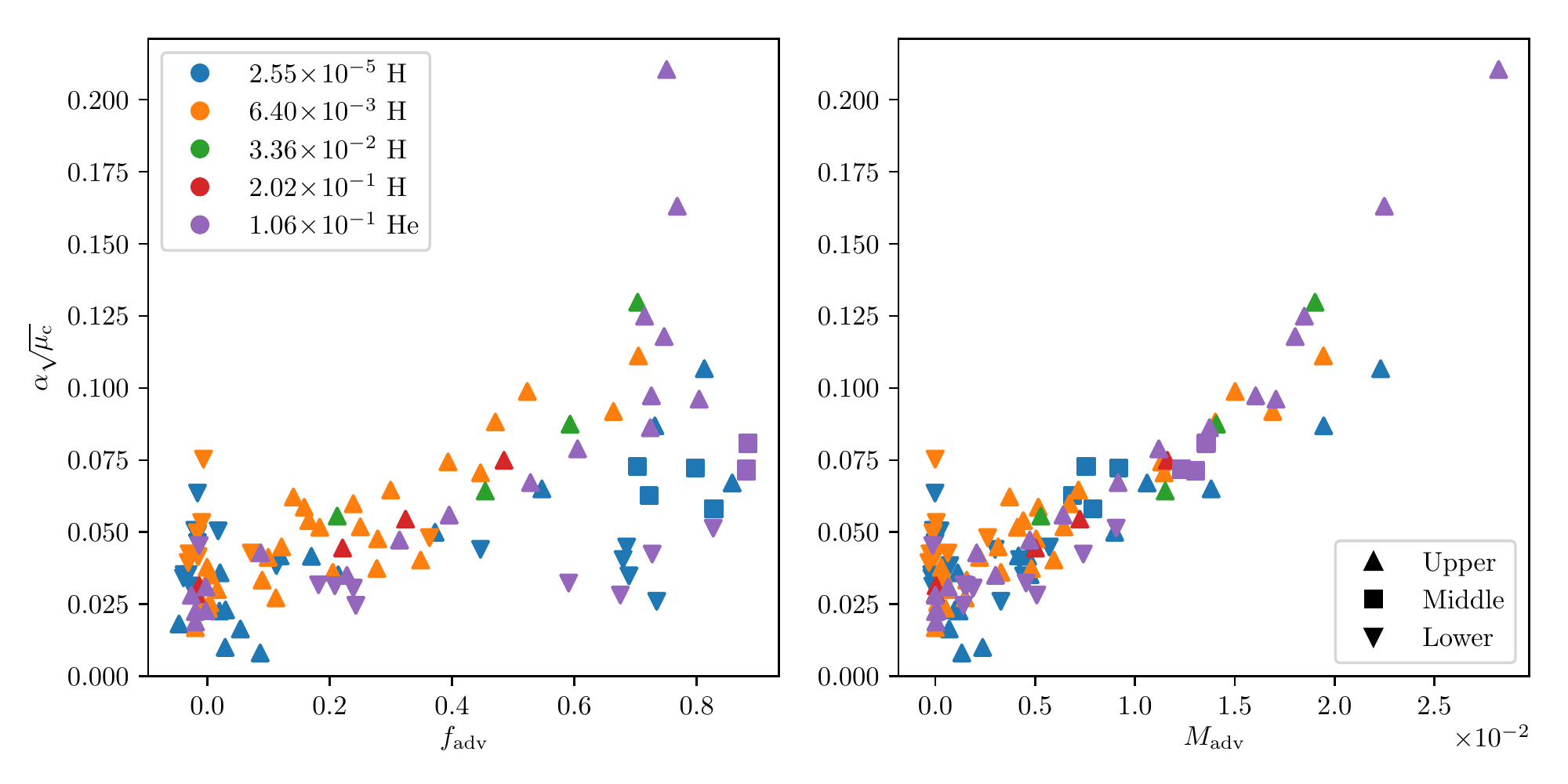}
	\caption{(Figure modified) Correlations of convective quantities with $\alpha\sqrt{\mu_{\rm c}}$, where $\mu_{\rm c}$ is the central/midplane mean molecular weight. The left panel shows the correlation with advective flux
		$f_{\rm adv}$ (see Eqn.~\ref{eqn:fadv}), while the right panel shows the correlation with advective Mach number $M_{\rm adv}$
		(see Eqn.~\ref{eqn:Madv}). The color of each symbol denotes the orbital frequency $\Omega$ (in rad s$^{-1}$) and the most abundant element
		(H or He) in the assumed chemical composition as listed in the upper left legend.
		The shape of the symbol corresponds to the branch to which the simulation settled, as listed in the lower right legend.
		Blue data comes from \citet{HIR15}, orange from \citet{HIR14}, green and red from \citet{COL16}, and purple from this work.
		The $\sqrt{\mu_{\rm c}}$ factor is solely used to tighten the correlation and in particular it corrects systematically lower values of $\alpha$ seen in the He simulations (for the same $f_{\rm adv}$ or $M_{\rm adv}$ as the H simulations). We also note that there are two outliers at $f_{\rm adv}\sim M_{\rm adv}\sim0$; these two simulations are not outliers in either $\alpha$ or $\mu_{\rm c}$ but have relatively high values for both of these with $\alpha\sim0.04-0.05$.
	}
	\label{fig:corr}
\end{figure*}

\begin{figure}
	\includegraphics[width=.95\linewidth]{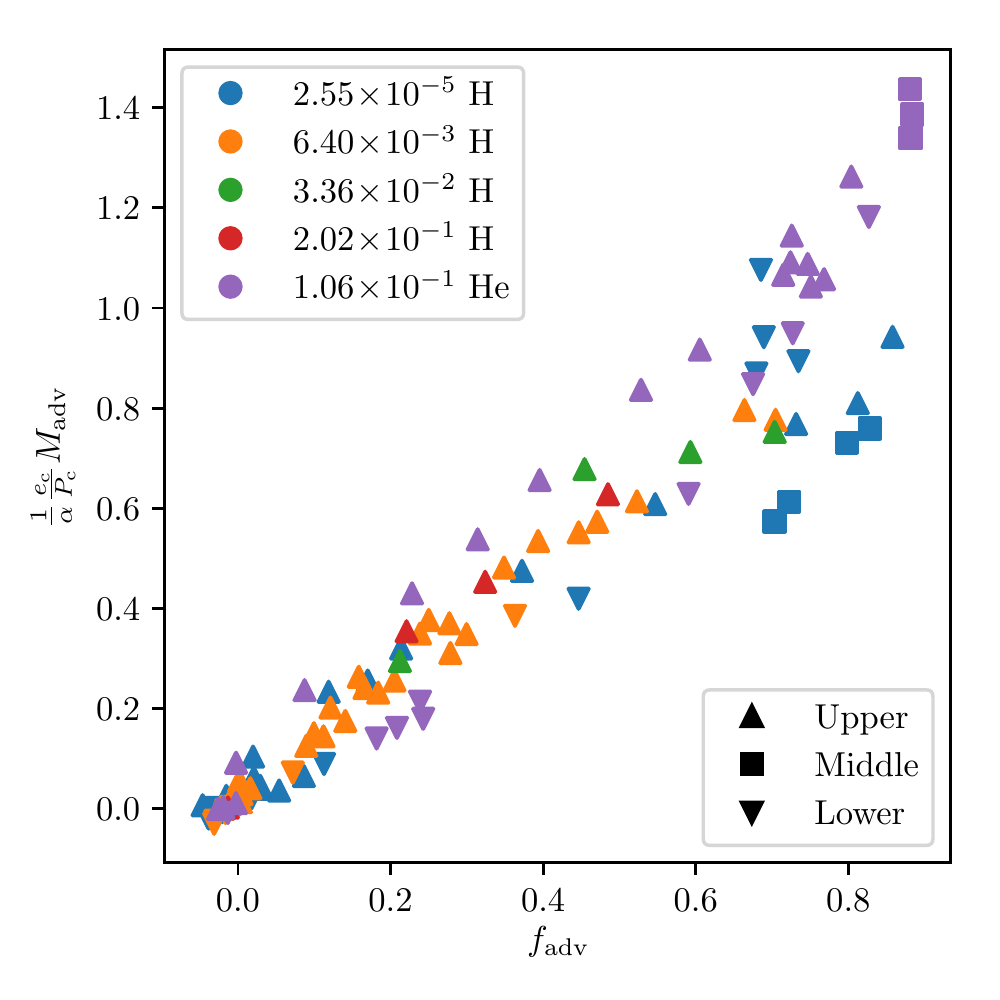}
	\vspace*{-1em}
	\caption{Correlations of  advective flux $f_{\rm adv}$ (see Eqn.~\ref{eqn:fadv}) with advective Mach number $M_{\rm adv}$ (see Eqn.~\ref{eqn:Madv}) times $e/(\alpha P)$.	The linear tend here shows that the relation \ref{eqn:f-M} is a reasonably good approximation.
	The color and shapes of the symbols are the same as those in Fig.~\ref{fig:corr} and each symbol represents time-averaged data from a single simulation. The spread at large $f_{\rm adv}$ in both this figure and in the left panel of Fig.~\ref{fig:corr} suggests that $M_{\rm adv}$ is the more relevant quantity to compare to $\alpha$.	
	}
	\label{fig:f-M}
\end{figure}

\subsection{Enhancement of $\alpha$}

As can be seen in Figure~\ref{fig:scurve}, our measured $\alpha$ values vary along the S-curve reaching a maximum of $\sim0.15$ near the tip of the upper branch, with low values of a few percent high up on the upper (outburst) branch and all along the lower (quiescent) branch. 
This gives a variation of $\alpha$ that is relatively large (a factor of $\sim 5$), similar to that inferred for DNe \citep[e.g.][]{SMAK99,KL12,HIR14,SCE18}, and consistent with modeling of AM CVn lightcurves, but notably higher than the inferred lower limit of the ratio of minimum and maximum $\alpha$ values \citep{KOT12}.

Following \citet{HIR14,HIR15}, we define the quantities $f_{\rm adv}$ as a means to estimate the fraction of vertical energy transport which is done by advection,
\begin{equation}
\label{eqn:fadv}
f_{\rm adv} \equiv \left\{\dfrac{\int\left\{\left<\left(e+E\right)v_z\right>\right\}_t{\rm sign}(z)\left<P_\text{th}\right>dz}{\int\left\{\left<F_{{\rm tot},z}\right>\right\}_t{\rm sign}(z)\left<P_\text{th}\right>dz}\right\}_t,
\end{equation}
and $M_{\rm adv}$ an estimate of the Mach number of advective eddies,
\begin{equation}
\label{eqn:Madv}
M_\text{adv} \equiv \left\{\dfrac{1}{\int\!\left<P_\text{th}\right>dz}\!\int\! \dfrac{\left\{\left<(e+E)v_z\right>{\rm sign}(z) \left<P_\text{th}\right>\right\}_t}{\left\{\left<(e+E)\right>\left<c_\text{s}\right>\right\}_t}dz\right\}_t\!,
\end{equation}
where $e$ is the gas internal energy density, $E$ is the radiation energy density, $v_z$ is the vertical velocity, $P_{\rm th}$ is the thermal pressure (gas plus radiation), and $F_{{\rm tot},z}$ is the total energy flux in the vertical direction, including Poynting
and radiation diffusion flux.

We first presented the results of convection enhancing $\alpha$ in \citet{HIR14}, where we showed that $f_{\rm adv}$ was correlated with $\alpha$ (see orange points in the left frame of Figure~\ref{fig:corr}). However, \citet{HIR15} noted that with additional data it became clear that $M_{\rm adv}$ was more correlated with $\alpha$, and \citet{SCE18} have also noted that $\alpha$ is not well correlated with $f_{\rm adv}$. Here we further confirm these results in Figure~\ref{fig:corr} where we present the data from this work along with those from \citet{HIR14}, \citet{HIR15}, and \citet{COL16}. With these data it seems clear that convection\footnote{We have yet to identify a way to conclusively distinguish convection from advection, although it is clear that our simulations are convectively unstable because they exhibit a negative vertical entropy gradient.} dominated energy transport through the disk is not sufficient to enhance $\alpha$. The convection must not be slow (i.e. $M_{\rm adv}\gtrsim 10^{-2}$) in order for it to enhance $\alpha$. 

From Equations \ref{eqn:fadv} and \ref{eqn:Madv} it is clear that $f_{\rm adv}$ and $M_{\rm adv}$ are closely related. By incorporating the \citet{SS73} prescription
\begin{equation}\label{eqn:SS}
F_{\rm tot}=\dfrac{F_{\rm adv}}{f_{\rm adv}}\sim \alpha P c_{\rm s}
\end{equation}
where $c_s$ is the sound speed, one can write the following relation\footnote{We thank the anonymous referee for pointing out this relation to us and for coming up with the concept of Fig.~\ref{fig:f-M}.}:
\begin{equation}\label{eqn:f-M}
f_{\rm adv}\sim\dfrac{M_{\rm adv}}{\alpha}\dfrac{e}{P}.
\end{equation}
In Fig.~\ref{fig:f-M} we show that this describes our data reasonably well. However, the spread in the relation increases with $f_{\rm adv}$, indicating that as convection saturates (i.e. as $f_{\rm adv}\rightarrow 1$) the connection between $f_{\rm adv}$ and $M_{\rm adv}$ breaks down. This may be related to the fact $e/P$ is not fixed in our simulations and the same physics (i.e. ionization transitions) which leads to high $f_{\rm adv}$ also causes $e/P$ to vary. For instance, in simulation $\Sigma$8.9e1-U1 $f_{\rm adv}\approx1$ and the midplane value of $e/P$ varies from $\sim5-6$. This combined with the fact that the correlations between $f_{\rm adv}$ and $\alpha\sqrt{\mu_{\rm c}}$ (Fig.~\ref{fig:corr}), and $f_{\rm adv}$ and $M_{\rm adv}$ (Fig.~\ref{fig:f-M}) worsen at roughly the same value of $f_{\rm adv}$ suggests that $M_{\rm adv}$ is the more relevant physical quantity to be compared with $\alpha$.

The correlation we find between $M_{\rm adv}$ and $\alpha$ is consistent with our hypothesis \citep{HIR14} that this enhancement of $\alpha$ is caused by convection advecting magnetic fields and seeding the MRI with vertical fields. The axisymmetric MRI then feeds off of this and generates stronger magnetic turbulence resulting in an enhancement in $\alpha$. However, we stress that the mean turbulent motion (i.e. the RMS value of $v_z$) is $\sim5-10$ times the speed of the convective eddies ($c_{\rm s}M_{\rm adv}$), making identifying the effect of convection on the MRI challenging. 
Despite $c_{\rm s}M_{\rm adv}$ being small compared to the mean vertical motion, we have demonstrated in our previous work that convective eddies are capable of advecting strong magnetic fields vertically (see \citealt{COL17}, in particular Figures 5 and 6). 

We also note there are many helium disk simulations (purple points) and \citet{HIR15} simulation (blue points) in the bottom right corner of the left panel of Fig.~\ref{fig:corr}. These points are almost all from the lower or middle branches, indicating that convection occurring at temperatures lower than the corresponding ionization transition is inefficient at enhancing $\alpha$. Both hydrogen and helium simulations occupy this parameter space, implying that low temperature convection gives low $\alpha$ for relatively high $f_{\rm adv}$ regardless of composition. 

Additionally, the helium simulations have systematically lower values for $\alpha$ compared to hydrogen simulations from the same branch with comparable $f_{\rm adv}$ or $M_{\rm adv}$. This seems to be related to the increased mean molecular weight as scaling $\alpha$ by $\sqrt{\mu_{\rm c}}$ (as was done in Fig.~\ref{fig:corr}) hides this effect and increases the correlation at $f_{\rm adv},\, M_{\rm adv} \gtrsim 0$, but worsens the correlation when convection is not present. It is not clear why $\mu$ is related to the convective enhancement of the MRI, however $\mu$ is known to be related to the efficiency of convection within the context of mixing-length theory \citep[e.g.][]{BV58,LUD02} suggesting that the two should be related.

\subsection{Persistent Convection}
\label{sec:persist}

\begin{figure}
	\vspace*{1em}
	\includegraphics[width=.95\linewidth]{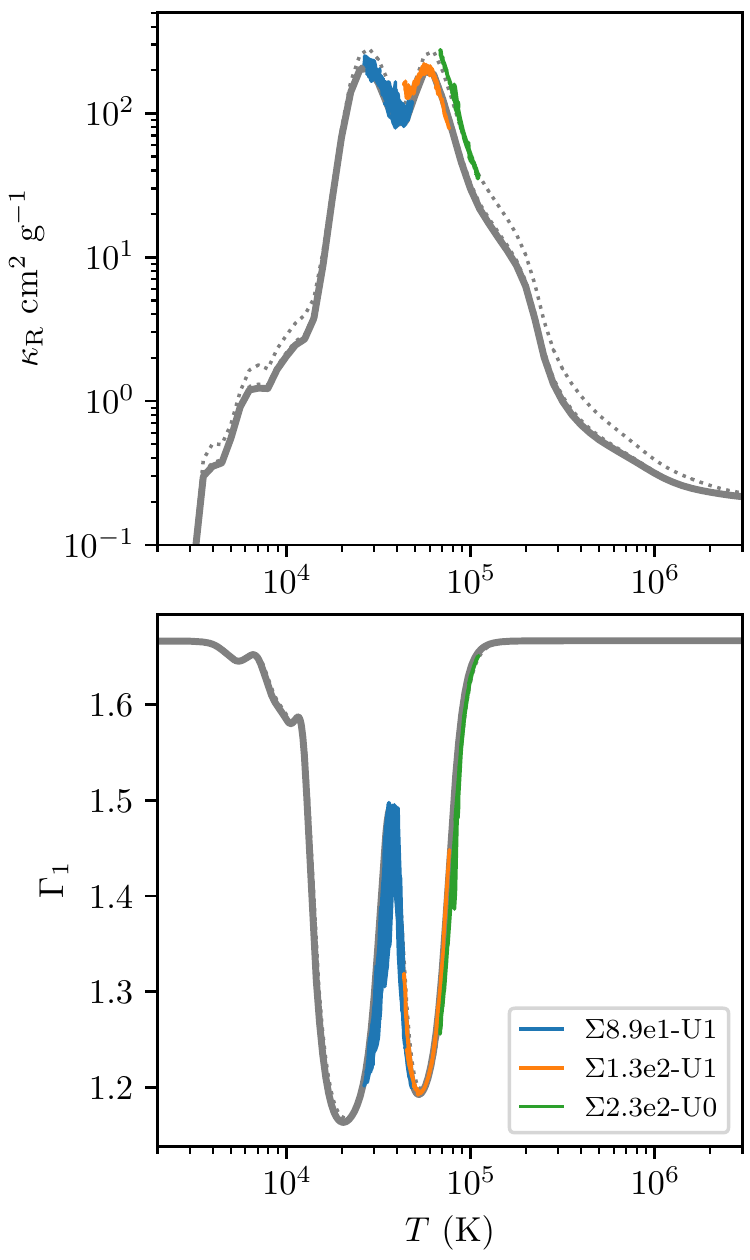}
	\caption{Rosseland mean opacity (top) and adiabatic index $\Gamma_1$ (bottom) as a function of temperature for three simulations: $\Sigma$8.9e1-U1 (blue), $\Sigma$1.3e2-U1 (orange) and $\Sigma$2.3e2-U0 (green). These simulations exhibit persistent convection, mostly persistent convection, and intermittent convection respectively and are also shown in the top three panels of Fig.~\ref{fig:By}. 
	Each gray curve corresponds to the $\kappa_{\rm R}$ and $\Gamma_1$ functions for a fixed density. 
	The solid gray line corresponds to the time average central density ($\bar{\rho_c}$) of simulation $\Sigma$8.9e1-U1 while the dotted lines correspond to $\bar{\rho_c}$ of the other two simulations (there is significant overlap of these curves).
	Each simulation is expected to evolve roughly along its corresponding gray curve.
	The colored lines correspond to the time evolution (from $t_1$ to $t_2$ listed in Table~\ref{table:param}) of the mean midplane values for each of the simulations. All of our AM CVn simulations exhibiting persistent convection spend most of their time within the concave region of the opacity function between the two peaks of the corresponding gray curve.
	}
	\label{fig:sim-k}
\end{figure}

One of the most noticeable differences compared to our previous work on DN simulations \citep{HIR14,COL16,COL17} is that our simulations of AM~CVn disks exhibit persistent convection. With the exception of one simulation (ws0837 from \citealt{HIR15}), all of our previous simulations exhibiting convection did so intermittently. 
\citet{SCE18} also found their DNe simulations to exhibit intermittent convection.
However, several of our AM CVn simulations exhibit persistent convection. In Fig.~\ref{fig:sim-k} we show the evolution of the midplane Rosseland mean opacity $\kappa_{\rm R}$ and adiabatic index $\Gamma_1$ for three simulations. These simulations are representative of the trend we find between persistence of convection and location in the opacity curve. Simulations which exhibit few to zero transitions out of epochs dominated by convective energy transport stay between the two maxima in Rosseland mean opacity resulting from the two ionizations of helium. This implies that the concave nature of the He opacity curve is what stabilizes the convective transport within the disk.  

\begin{figure}
	\includegraphics[width=\linewidth]{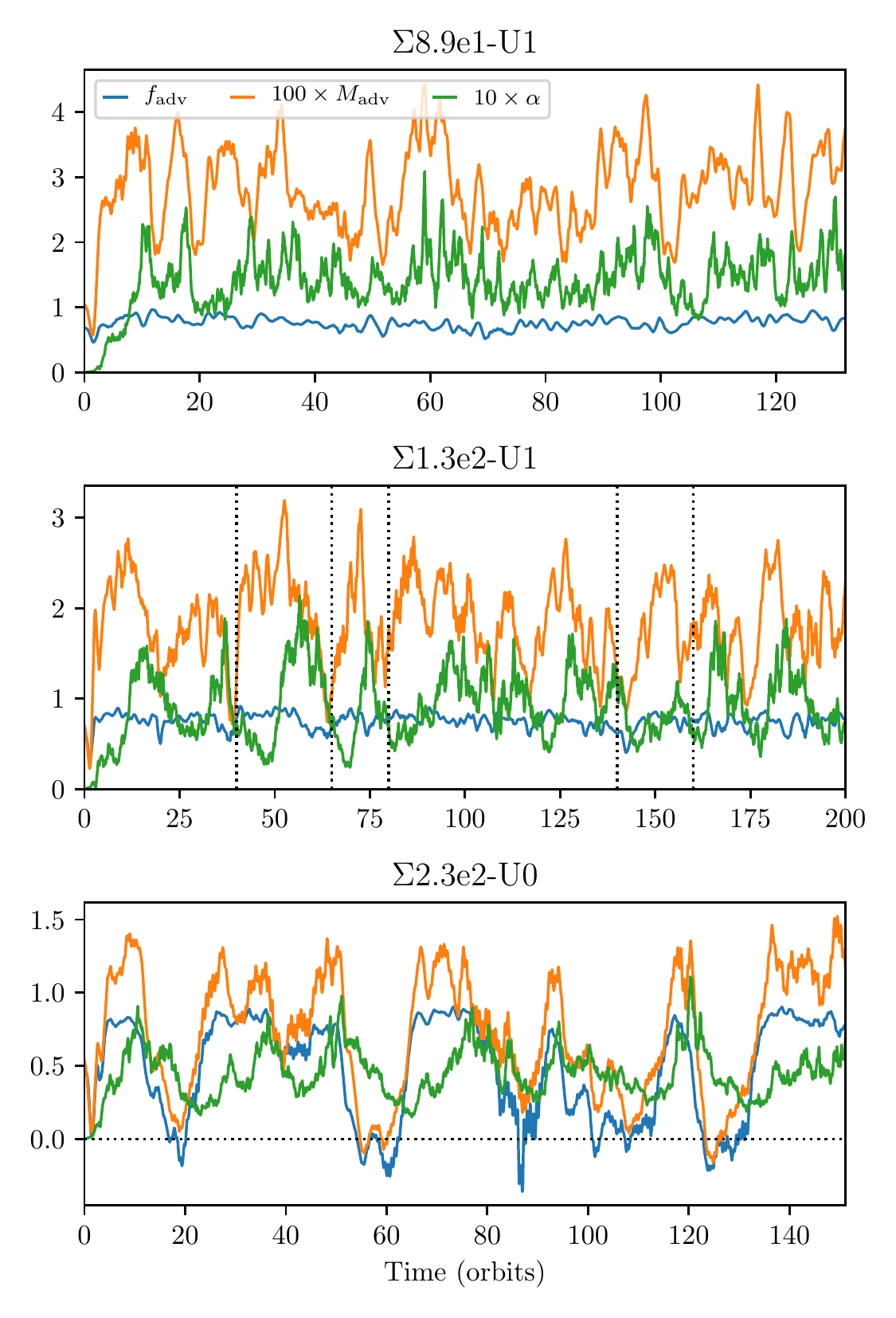}
	\caption{
	$f_{\rm adv}$, $M_{\rm adv}$, and $\alpha$ as a function of time for three simulations: $\Sigma$8.9e1-U1 (top), $\Sigma$1.3e2-U1 (middle) and $\Sigma$2.3e2-U0 (bottom). $M_{\rm adv}$, and $\alpha$ are multiplied by factors of 100 and 10 respectively. The vertical dotted lines for $\Sigma$1.3e2-U1 denote times where the magnetic field structure changes (see Fig.~\ref{fig:By}). Negative values of $f_{\rm adv}$ and $M_{\rm adv}$ are indicative of inward advection of energy. These values can temporarily have different signs due to slightly different smoothing and averaging procedures.
	}
	\label{fig:time}
\end{figure}

\begin{figure*}
	\includegraphics[width=1.0\linewidth]{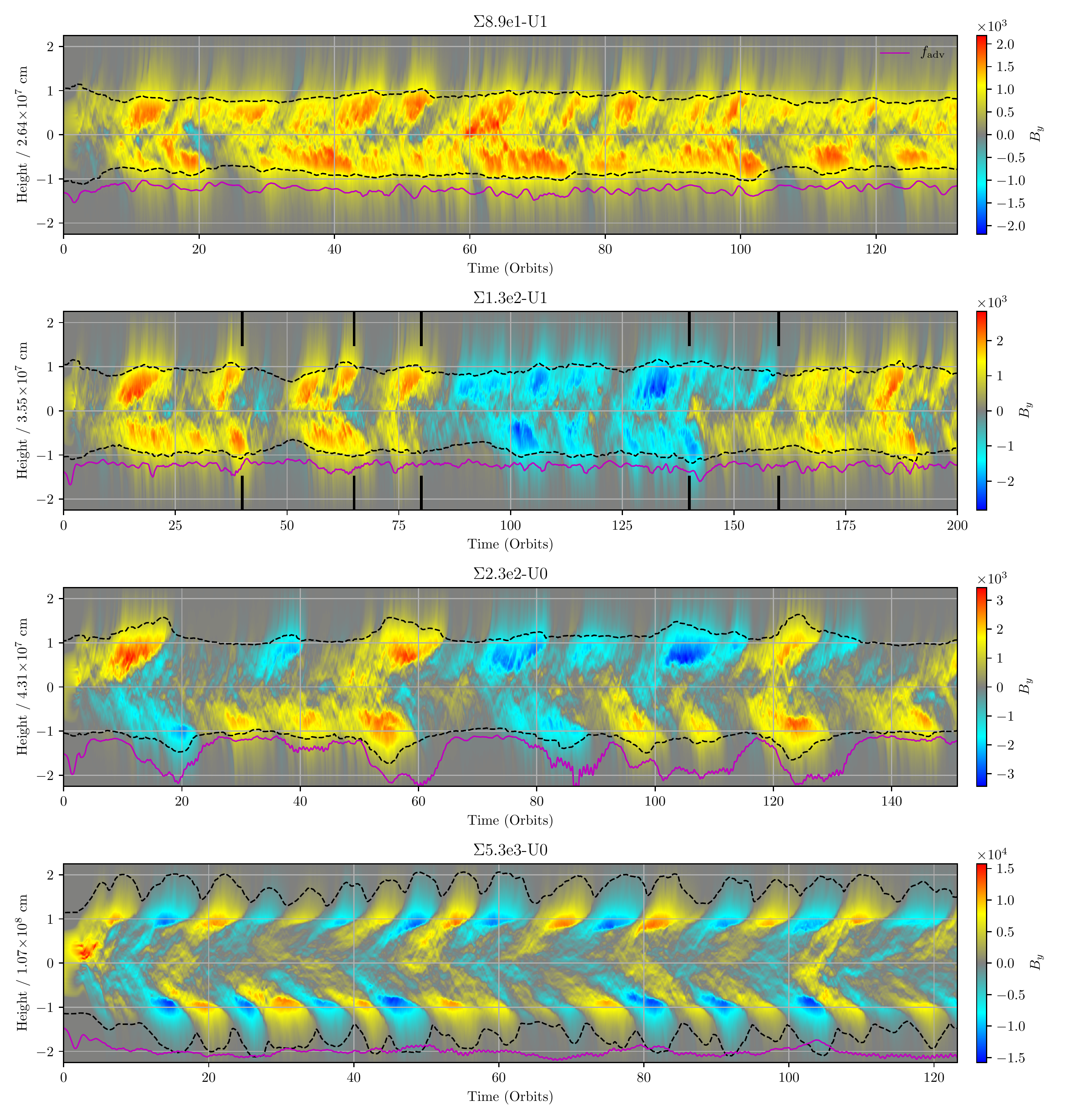}
	\caption{
		Horizontally-averaged azimuthal magnetic field $\have{B_y}$ as a function
		of time and height for simulations which exhibit varying levels of convection.
		The normalizations indicated in the vertical axes are the respective
		simulation length units.
		The dashed black lines show the time-dependent heights of
		the photospheres in the horizontally averaged structures.
		The convective fraction $f_{\rm adv}-2$ (see Eqn. \ref{eqn:fadv}) is plotted in magenta to highlight the connection between the dynamo and convection. Note that $f_{\rm adv}-2$ uses the same vertical scale as $B_y$, i.e. when the magenta line is near $-1$ then $f_{\rm adv}\approx 1$.
		Focusing on $B_y$, the non-convective simulation $\Sigma$5.3e3-U0 (bottom panel) shows the standard pattern of field reversals normally associated with the butterfly diagram. 
		In the simulation which exhibits intermittent convection ($\Sigma$2.3e2-U0) where $f_{\rm adv}$ is high, the field tends to maintain its sign and changes in sign/parity are often associated with a dip in $f_{\rm adv}$. Of the two persistently convective simulations displayed here, $\Sigma$8.9e1-U1 shows no global field reversals, and $\Sigma$1.3e2-U1 has a few reversal events which seem uncorrelated with $f_{\rm adv}$, however the marked times for this simulation which correspond to field reversals also correspond to dips in $M_{\rm adv}$ (see Fig.~\ref{fig:time}).
		Note that the simulations from the top three panels are also shown in 
		Figures \ref{fig:sim-k} and \ref{fig:time}.
	}
	\label{fig:By}
\end{figure*}

The simulation $\Sigma$2.3e2-U0 (green curves in Fig.~\ref{fig:sim-k} and third panel in Figs.~\ref{fig:time} and \ref{fig:By}) exhibits intermittent convection similar to what we presented in Section 3.4 of \citet{HIR14}. We originally described this behavior as a limit cycle, which we summarize here. Simulations towards the tip of the upper branch often switch between epochs where energy transport is dominated by radiative diffusion and convection. During convective epochs, turbulent stress and dissipation is increased causing the disk to heat. This heating combined with large and negative values of $\dd\kappa_{\rm R}/\dd T$ makes it easier for photons to escape and the disk transitions to a radiative epoch. This transition is accompanied by a decline in turbulent heating, causing the disk to be slightly too effective at removing heat. This leads to cooling and higher opacities which then initiates convection, thereby completing the limit cycle. 

From this description it is clear that the negative values of $\dd\kappa_{\rm R}/\dd T$ on the upper branch of the S-curve are fundamental to driving this limit cycle. The double peaked nature of the He opacity curve ensures that the opacity increases regardless of heating/cooling, preventing the disk from losing optical depth, thereby preventing a transition to radiative epochs.
In other words, the increase in opacity in both directions always requires superadiabatic temperature gradients to get
the heat out.

This finding also shows that the enhancement of $\alpha$ does not require convection to be intermittent, causing our emphasis on the convective limit cycle in Section 3.4 of \citet{HIR14} to be slightly misleading. While this limit-cycle is interesting and has given us some insight into the dynamo \citep{COL17}, it does not seem to play a key role in enhancing $\alpha$. 
This can be visualized\footnote{We thank the referee for encouraging us to add this plot.} in Fig.~\ref{fig:time} which shows that $\alpha\gtrsim 0.1$ and $f_{\rm adv}\approx 1$ for the duration of simulation $\Sigma$8.9e1-U1. Although, this need not be the case as $\Sigma$1.3e2-U1 maintains $f_{\rm adv}\approx 1$ but has dips in $\alpha$ which seem to lag dips in $M_{\rm adv}$ by $\sim 10$ orbits. Finally, $\Sigma$2.3e2-U0 shows intermittent convection similar to that in \citet{HIR14} where dips in $\alpha$ and $f_{\rm adv}$ appear to be correlated.

\subsection{Quenching of Dynamo Reversals}

In the simulations presented here, we found that convection modifies the standard dynamo found in accretion disk simulations \citep[see e.g.][]{BRA95,DAV10} by quenching magnetic field reversals (see Figure~\ref{fig:By}). We discussed this particular phenomenon in depth in \citet{COL17}. We show the dynamo behavior of four of our simulations in Figure~\ref{fig:By} with persistent, nearly-persistent, intermittent, and no convection.

Our simulations with persistent convection offer a new way to examine this quenching over longer timescales. For the case of our persistent simulations, the dynamo maintains a constant sign of $B_y$ for long durations. In \citet{COL17} we hypothesized that the polarity of the magnetic fields at the beginning of a convective epoch is held fixed until the simulation becomes convectively stable. This is consistent with the fact that our persistently convective simulations tend to maintain the same parity and sign of $B_y$ as that of our initial conditions\footnote{This indicates that to perform a proper dynamo study of convective accretion disks, it is necessary to examine multiple initial field configurations. However, this is beyond the scope of this work.}. 
However, we do see some magnetic field reversals in persistently convective simulations which seem uncorrelated to $f_{\rm adv}$. In these cases there is evidence that dips in $M_{\rm adv}$ are associated with field reversals while $f_{\rm adv}\sim 1$ (see marked times in Figs. \ref{fig:time} and \ref{fig:By} for simulation $\Sigma$1.3e2-U1).

\subsection{Validity of Ideal MHD}
\label{sec:ideal}

For DNe (with a hydrogen dominated composition) it is unclear how well the plasma within the disk can be described by MHD \citep[see e.g.][]{GAM98}. As we discussed in \citet{COL16}, post-processing of our DNe simulations (which assume ideal MHD) suggest that non-ideal MHD effects are important along the entire quiescent branch. In particular we found that Ohmic dissipation is important, the Hall term may need to be taken into consideration, and that ambipolar-diffusion was negligible. \citet{SCE18} incorporated Ohmic dissipation directly in their simulations, neglecting the Hall term,  and found that
the transition from ideal to non-ideal MHD occurred below the tip of quiescent branch. They estimated that the Hall term is an order of magnitude smaller than Ohmic dissipation, leading them to conclude that only Ohmic dissipation is important. In contrast to these results for DNe, we find here that ideal MHD should always be applicable to AM CVns.  The reason for this is that the abundant elements C, N, and O remain ionized in the quiescent state, and are therefore copious sources of free electrons.

\begin{figure}
	\includegraphics[width=1.0\linewidth]{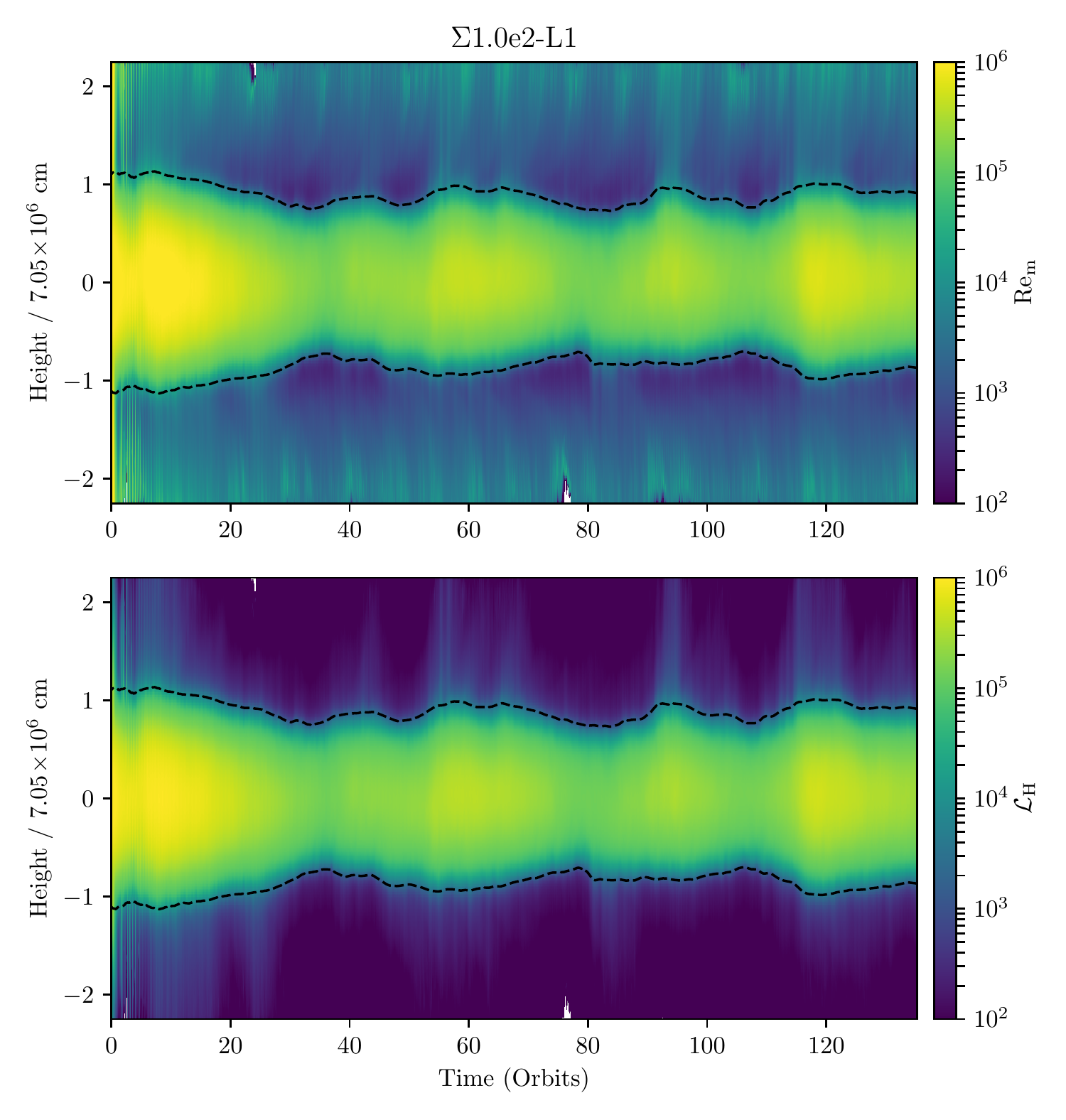}
	\caption{
		Horizontally averaged magnetic Reynolds and Hall Lundquist numbers for the coldest optically thick lower branch simulation ($\Sigma$1.0e2-L1). Lower values corresponds to non-ideal MHD effects being more important. Re$_{\rm m}$ (Eq. \ref{eqn:Rm}, top panel) corresponds to Ohmic dissipation and $\mathcal{L}_{\rm H}$ (Eq. \ref{eqn:Ha}, bottom panel) corresponds to the Hall term. The black dashed contour is the photosphere. The large values of Re$_{\rm m}$ and $\mathcal{L}_{\rm H}$ that we recover are constant with the approximation of ideal MHD \citep[see e.g.][]{HAW96,SAN02,SCE18}.
	}
	\label{fig:nonideal}
\end{figure}

To estimate the validity of ideal MHD we computed the magnetic Reynolds number (Re$_{\rm m}$) following \citet{FLE00} and the Hall Lundquist number ($\Lambda_{\rm H}$) following \citet{SCE18} assuming thermal ionization from our AM CVn simulation data (see Fig~\ref{fig:nonideal}).
\begin{align}
\label{eqn:Rm}
{\rm Re}_{\rm m}\equiv\dfrac{c_{\rm s}^2}{\Omega \eta_{\rm O}},
\end{align}
where $\eta_{\rm O}=230T^{1/2}n_n/n_e$~cm$^2$~s$^{-1}$ is the Ohmic diffusivity coefficient \citep{BLA94}, $n_n$ and $n_e$ are the number density of the neutrals and electrons respectively.
\begin{align}
\label{eqn:Ha}
\mathcal{L}_{\rm H}\equiv\sqrt{\dfrac{4\pi}{\rho}}\dfrac{n_eeH_P}{c},
\end{align}
where $\rho$ is mass density, $e$ is the elementary charge, $H_P$ is the pressure scale height, and $c$ is the speed of light. For our coldest optically thick simulation both of these numbers are $\gtrsim 10^4$ within the photosphere. This combined with previous studies of non-ideal MHD \citep[e.g.][]{HAW96,SAN02,SCE18} suggests that ideal MHD is a good approximation for AM CVns, even throughout quiescence.

\section{Discussion}
\subsection{Observational Constraints}

\begin{figure}
	\includegraphics[width=1.0\linewidth]{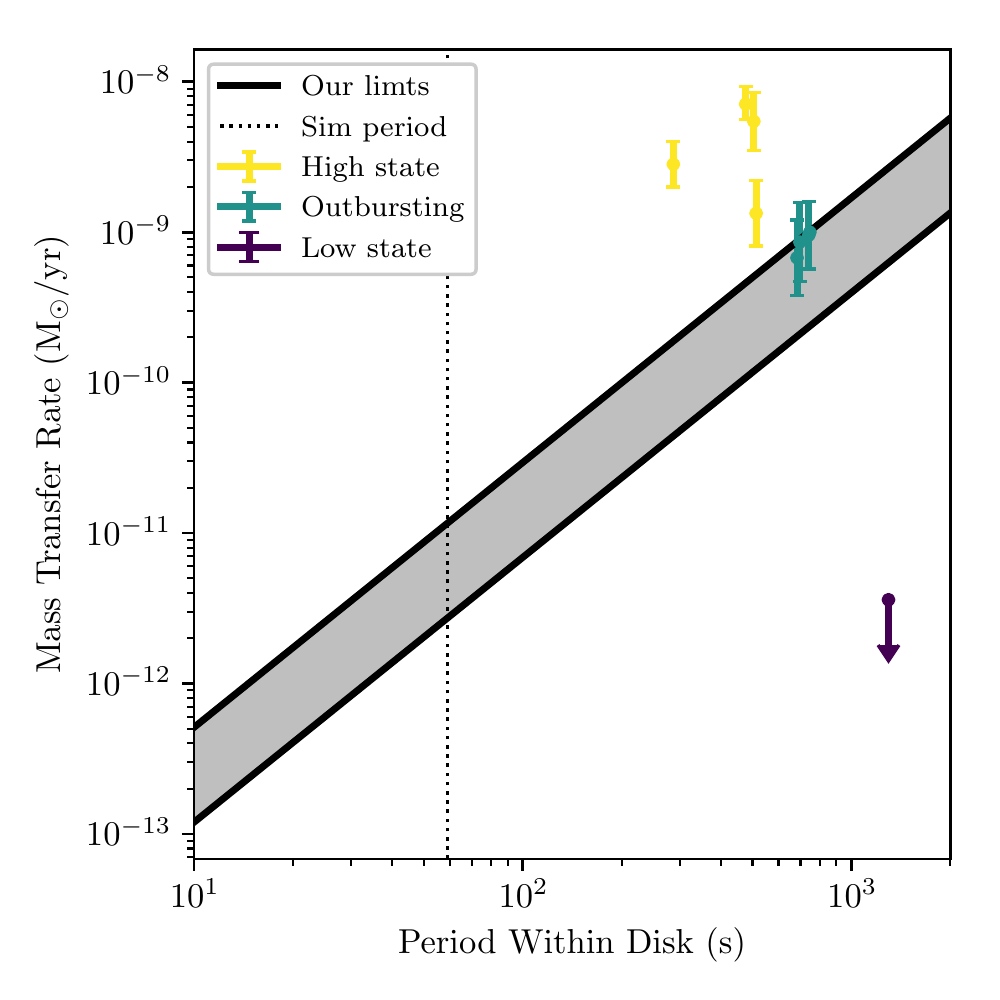}
	\caption{
		Limiting mass transfer rates for local instability within AM CVn accretion disks as a function of orbital period. Solid black lines are limiting transfer rates inferred from our simulation S-curve (Fig.~\ref{fig:scurve}) which was computed at an orbital period of $59$ s (vertical dotted black line). The region between these lines is shaded gray and denotes the presence of a local instability within the accretion disk; how this local instability relates to the onset of outbursts (a global instability) is not entirely clear. Our lines for limiting transfer rates are extrapolated from this orbital period assuming Eqns.~\ref{eqn:balance} and \ref{eqn:TP}.  
		We also plot the data for several observed AM CVn with observationally inferred mass transfer rates (see Table~\ref{table:amcvn}) assuming that the orbital period corresponding to the outer edge of the accretion disk is $0.6^{1.5}$ times the orbital period of the binary. The coloring of these points corresponds to the observed state of the AM CVns with persistent high state systems in yellow, outbursting systems in green, and the persistent low state system in purple (this system only has an observational upper bound on the transfer rate). Note that if the outer edge of an AM CVn disk was found to lie just below our $\dot{M}^-$ line this would imply that a significant fraction of the disk is unstable and would likely still exhibit outbursts.
	}
	\label{fig:obs}
\end{figure}

\begin{deluxetable*}{ccccccc}
	\tablecaption{Observational AM CVn data\label{table:amcvn}}
	\tablehead{
		 & \colhead{$P_{\rm orb}$} & \colhead{$P_{\rm disk}$} & \colhead{$\dot{M}_{\rm low}$} & \colhead{$\dot{M}$} & \colhead{$\dot{M}_{\rm high}$} & \\
		\colhead{System Name} & \colhead{(min)} & \colhead{(min)} & \colhead{$(10^{-9} M_\sun \,{\rm yr}^{-1})$} & \colhead{$(10^{-9} M_\sun \,{\rm yr}^{-1})$} & \colhead{$(10^{-9} M_\sun \,{\rm yr}^{-1})$} & \colhead{Source}
	}
	\startdata
\hline
\multicolumn{7}{c}{Persistent High State}\\
\hline
ES Cet & 10.3 & 4.79 & 2 & $2.83^*$ & 4 & \citet{ESP05} \\
AM CVn & 17.1 & 7.95 & 5.6 & $7.1$ & 9.3 & \citet{ROE07} \\
KIC 004547333 (SDSS J1908) & 18.1 & 8.41 & 3.5 & $5.45^*$ & 8.5 & \citet{FON11} \\
HP Lib & 18.4 & 8.55 & 0.81 & $1.33^*$ & 2.2 & \citet{ROE07} \\
\hline
\multicolumn{7}{c}{Outbursting}\\
\hline
CR Boo & 24.5 & 11.4 & 0.38 & $0.675^*$ & 1.2 & \citet{ROE07} \\
KL Dra & 25 & 11.6 & 0.47 & $0.859^*$ & 1.57 & \citet{RAM10} \\
V803 Cen & 26.6 & 12.4 & 0.57 & $0.955^*$ & 1.6 & \citet{ROE07} \\
\hline
\multicolumn{7}{c}{Persistent Quiescence}\\
\hline
GP Com & 46.5 & 21.6 &  &  & 0.0036 & \citet{ROE07}\\
\enddata
\tablecomments{
$P_{\rm orb}$ is the orbital period of the binary, $P_{\rm disk}$ is the estimated orbital period at the outer disk edge (see Eqn.~\ref{eqn:Pdisk}), $\dot{M}_{\rm low}$ and $\dot{M}_{\rm high}$ are the lower and upper bounds on the mass transfer rate respectively. $\dot{M}$ is the estimated mass transfer rate; values denoted by an asterisk are simply taken to be the geometric mean of the upper and lower bounds.}
\end{deluxetable*}

Based on our S-curve for $\Omega=0.106$~rad s$^{-1}$ (Figure~\ref{fig:scurve}) our data suggest that the critical effective temperatures $\Tep\approx11500$ K
and $\Tem\approx8000$ K at the annulus we simulated. For the discussion that follows we assume that torques on the disk by the binary companion and the accreting white dwarf and other edge effects are negligible, therefore
\begin{equation}\label{eqn:balance}
\sigma T_{\rm eff}^4 = \dfrac{3}{8\pi}\dot{M}\Omega^2 = \dfrac{3\pi}{2}\dot{M}P_{\rm acc}^{-2},
\end{equation}
where $P_{\rm acc}=2\pi/\Omega$ is the orbital period for a given annulus in the accretion disk.
This implies our simulated annulus is unstable to mass accretion rates between $\sim 3\!\times\!10^{-12}$ and $\sim 10^{-11}\,M_\sun$ yr$^{-1}$. To understand instability criteria for normal AM CVn outbursts we need \Tep and \Tem for the whole disk. Since our data is limited to one annulus we assume
\begin{equation}\label{eqn:TP}
T_{\rm eff}^{\pm}\propto P_{\rm acc}^{-.06},
\end{equation}
based on the scalings from Eqns. 7 and 9 of \citet{KOT12}. From this we derive the following limiting mass transfer rates as a function of orbital period within the accretion disk:
\begin{align}
\label{eqn:M+}
\dot{M}^+&=1.2 \times 10^{-11}\,{\rm M}_\sun\,{\rm yr}^{-1} \left(\dfrac{P_{\rm acc}}{1\,{\rm min}}\right)^{1.76}\\
\label{eqn:M-}
\dot{M}^-&=2.8 \times 10^{-12}\,{\rm M}_\sun\,{\rm yr}^{-1} \left(\dfrac{P_{\rm acc}}{1\,{\rm min}}\right)^{1.76}.
\end{align}
We stress that these limits are for a local instability at a given annulus (in terms of its orbital period) within an accretion disk. For a global instability (i.e. an outburst) to occur, presumably a significant range of annuli must be unstable. In other words instability at a single annulus is a necessary, but not sufficient condition for the onset of outbursts.

To compare these results to observed AM CVns we need estimates for the mass transfer rate and outer disk edge. The former can be inferred from parallaxes and luminosities \citep[see e.g.][]{ROE07}.
As for the later, the outer disk edge ($R_{\rm d}$) can be estimated as \citep{WAR03}
\begin{equation}
\dfrac{R_{\rm d}}{a}=\dfrac{0.6}{1+q},
\end{equation}
where $a$ is the semi-major axis of the binary and $q\leq1$ is the mass ratio.
Assuming that $q$ is small gives us a rough estimate of the orbital period of the outer disk edge ($P_{\rm disk}$) in terms of the orbital period of the binary ($P_{\rm orb}$):
\begin{equation}\label{eqn:Pdisk}
P_{\rm disk}\approx 0.6^{1.5} P_{\rm orb}=0.465\,P_{\rm orb}.
\end{equation}
We compiled observational data for several AM CVns in Table~\ref{table:amcvn} and plotted this data along with our critical mass transfer rates in Fig.~\ref{fig:obs}. This shows that our simulations are consistent with observed AM CVns.

\subsubsection{Lightcurves}

Another observational aspect that could be tested is the appearance of outburst lightcurves, as we did for DNe in \citet{COL16}. In that work all of our lightcurves based on our MHD simulations had peculiar zig-zagging decays from outburst called reflares, which are not observed in standard DNe. Here we speculate on how the issue of reflares might change for the AM CVn case. However, we leave the computation of outburst lightcurves for another paper.

Both the variation of alpha and the locations of the ends of the upper and lower branches play significant roles in shaping the lightcurve generated by the associated disk instability, as noted in our previous work in \citet{COL16}. The contrast in the critical $\Sigma$ values (ends of the branches) and the variation of $\alpha$ is comparable to what we found for the DN case, suggesting that reflares may be an issue for utilizing the disk instability model (DIM) to generate light curves for AM CVns.

Despite this, there are some aspects of our AM CVn simulations which could alleviate the reflare problem.
We note that for our DN work, the physical assumptions (e.g. high optical depth, ideal MHD) made in both our {\sc zeus} simulations and DIM models start to break down at the end of the lower branch, and these two methods also disagree on the end of the lower branch. If these issues are the cause of the reflares found in \citet{COL16}, then re-performing these calculations for the AM CVn case could result in lightcurves free of reflares, as both ideal MHD and high optical depths are good approximations. Also, because the convection towards the end of the upper branch is persistent (instead of intermittent) for AM CVns, using temporal means of the simulations as inputs for the DIM may result in better agreement between the two.

Additionally, observations of lightcurves for AM CVns pale in comparison to DNe. This is a reflection of the scarcity of these sources more than anything else; there are only $\sim 30$ known AM CVns with $\sim 1/3$ of them outbursting with normal outburst duration $\sim 1$ day \citep{KOT12,LEV15}. In contrast, the Catalina Real-time Transient Survey alone has over 700 classified DNe \citep{COP16}. This makes the observational constraints of AM CVn normal light curves, including lack of reflares, much weaker. Therefore the addition of high cadence high quality lightcurves of normal outbursts in AM CVns would enable better tests of accretion disk theory.

\subsection{Compositional Dependence}

We found that the convective enhancement of the MRI (measured through $\alpha$) is weaker for higher mean molecular weight at a given $f_{\rm adv}$ or $M_{\rm adv}$; this is demonstrated in Fig.~\ref{fig:corr} by scaling $\alpha$ by $\sqrt{\mu_{\rm c}}$.
As discussed in Section \ref{sec:persist} we also found the stability/persistence of convection is significantly different between disks which are H dominated (e.g. DNe) and He dominated (e.g. AM CVns), with convection in He disks exhibiting persistent convection. Finally, we found that the compositional difference between DNe and AM CVns results in significantly different conductivities in quiescence for these systems (see Section~\ref{sec:ideal}). This is because even when He is neutral the disk is still sufficiently hot for C, N, and O to be singly ionized enabling high conductivity throughout AM CVn quiescence, impling that the difference in $\alpha$ between the hot and cold states cannot be explained by varying conductivities for AM CVns.
These remarkable differences caused by composition are only identifiable through proper treatments of thermodynamics and radiation. While it remains to be seen how accurately our simulations represent nature, it is clear that this microphysics significantly affects the outcome and it would be na\"ive to believe that nature is indifferent to this.

\section{Conclusions}

In this paper we presented radiation-MHD stratified shearing-box simulation of AM CVn accretion disks. We found many similarities to our previous work on H dominated DNe accretion disks \citep[e.g.][]{HIR14}. In particular we found a similar S-curve (Fig.~\ref{fig:scurve}) and trend in the variation of $\alpha$, with enhancement only occurring near the tip of the upper branch, with peak values at the tip of $\alpha\sim 0.15$. This suggests that normal outbursts in AM CVns are very similar to those found in DNe. Likewise we also found that this enhancement of $\alpha$ is caused by convection which only manifests itself near the tip of the upper branch due the sharp temperature dependence of the EOS and opacities (see Figs. \ref{fig:eos}, \ref{fig:opacity}, and \ref{fig:sim-k}). Additionally, this convection acts to quench magnetic field reversals (see Fig. \ref{fig:By}) as we previously found in \citet{COL17}.

While these two types of accretion disks have many similarities, we noted some significant differences between our AM CVn and DN simulations: 1) Convective enhancement of $\alpha$ and the mean molecular weight are anticorrelated. 2) AM CVn simulations which lie between the two He opacity peaks exhibit persistent convection, while the convection found in H dominated disks is almost always intermittent. 3) The high ionization temperature of neutral He enables C, N, and O to be ionized even when He is neutral, leading to non-ideal MHD terms being negligibly small throughout quiescence in AM CVns. Contrastingly, non-ideal MHD terms are suspected to become important for DNe somewhere on the quiescent branch \citep[see e.g.][]{GAM98,COL16}. The uncertainty of where this transition occurs can be highlighted by the fact that \citet{HIR14} and \citet{SCE18} recover different endpoints for the lower branch\footnote{Although it should be noted that these works perform their analysis at slightly different $\Omega$, which is known to effect the S-curve.}, with \citealt{HIR14} reporting lower $T_{\rm eff}$ and higher $\Sigma$ for the tip of the lower branch. This causes them to disagree on where the transition away from ideal MHD lies. 
However, the value of $T_{\rm eff}$ where \citet{SCE18} find non-ideal effects become important is consistent with the results of \citet{HIR14} and their follow-up work \citet{COL16}.

We also compared our simulation inferred instability criteria to observations (see Fig.~\ref{fig:obs} and Eqns. \ref{eqn:M+} and \ref{eqn:M-}), finding these to be consistent with each other, further signifying the success of convectively enhanced MRI in describing nature.

\section*{Acknowledgements}

We thank the anonymous referee for a constructive report that led to
significant improvements in this paper.
We also thank Jared Brooks, Evan Bauer, Paul Groot, Iwona Kotko, and Jean-Pierre Lasota for their useful discussions and insight generated from their work. This research was supported by the United States National Science Foundation (NSF) under grant AST-1412417. Several of the above discussions were facilitated by a KITP program and so this research was supported in part by the NSF Grant PHY-1125915.
This work used the Extreme Science and Engineering Discovery Environment (XSEDE), which is supported by National Science Foundation grant number ACI-1548562. Specifically, we utilized the {\sc Comet} cluster at the San Diego Supercomputer Center at UC San Diego through allocation TG-AST160025.
The EOS and opacity calculations were
partly carried out on the Cray XT4 at CfCA, National Astronomical
Observatory of Japan, and on SR16000 at YITP in Kyoto University.
MC gratefully acknowledges support from the Institute for Advanced
Study, NSF via grant AST-1515763, and NASA via grant
14-ATP14-0059.
SH was supported by Japan JSPS KAKENHI 15K05040 and the joint research
project of ILE, Osaka University.

\software{{\sc zeus} \citep{Zeus1,Zeus2,ZeusRad}, {\sc phoenix} \citep{FER05}}
\bibliographystyle{aasjournal}
\bibliography{citations}

\label{lastpage}

\end{document}